# Voltage-Controlled Reversible Modulation of Colloidal Quantum Dot Thin Film Photoluminescence

*Sihan Xie‡, Han Zhu‡, Melissa Li, Vladimir Bulović\**

Research Laboratory of Electronics, Massachusetts Institute of Technology, Cambridge, Massachusetts 02139, United States

‡These authors contributed equally to this work.
*Corresponding author. Email: bulovic@mit.edu

**Abstract**

Active modulation of quantum dot thin film photoluminescence (PL) has far-reaching potential applications in biomedical and optoelectronic systems, but challenges remain in achieving large PL modulation depth and fast temporal response. Here we report an efficient voltage-controlled optical down-converter by optically exciting a colloidal quantum dot thin film within a quantum dot light-emitting diode (QD-LED) under reverse bias. Utilizing field-induced luminescence quenching, we show that a large electric field can strongly modify carrier dynamics in this nanostructured device, resulting in stable and reversible photoluminescence quenching. The device exhibits photoluminescence reduction of up to 99.5%, corresponding to a contrast ratio of 200:1, under the applied electric field of 3 MV/cm, with a 300 nanosecond response time. Using excitation wavelength dependent and transient PL spectroscopy, we further show that the high degree of quenching is achieved by a synergistic interplay of quantum-confined Stark effect (QCSE) and field-induced exciton dissociation.

Keywords: quantum dot, light-emitting diode, photoluminescence, electric field, quantum confined stark effect, exciton lifetime

Colloidal quantum dots (QDs) are a class of luminophores that enable applications ranging from biomedical sensing to next generation display and lighting technologies.[1–4] Substantial progress has been made in development of electrically driven QD light-emitting diodes (QD-LEDs).[5–7] However, challenges such as poor device stability,[8,9] efficiency droop at high brightness,[10] and large-area solution-processed patterning of QD films[11,12] remain as hurdles toward commercialization of QD-LEDs. One central problem that limits the electrically-driven device efficiency is the QD photoluminescence quenching due to either charge accumulation at or near the QD sites[13,14] or the increased electric field across the QD layer.[10,15] QD charging can efficiently quench QD PL by activating Auger non-radiative recombination pathways,[16,17] which involve the annihilation of an excited electron-hole pair by coupling its energy to an extra electron or a hole on the QD. Furthermore, an external electric field can polarize a QD exciton by physically separating the electron-hole pair in opposite directions.[18] The reduction in wavefunctions overlap lowers the radiative recombination rate and quenches QD PL, and is accompanied by a concomitant red-shift of the PL spectrum, known as the quantum-confined Stark effect (QCSE)[19]. At large electric fields, the columbic binding of excitons can be overcome, resulting in dissociation and formation of free carriers.[20]

Instead of modulating light emission by forming excitons through electrical charge injection, we can leverage the electric-field-induced QD PL quenching processes to control the PL efficiency of an optically excited QD film. Prior to our work, active modulation of QD PL in solid state has been demonstrated in a variety of structures, yet the overarching goals of achieving large PL quenching and a fast temporal response remained an active area of research. Bozyigit *et al.* used a QD capacitor structure to show that electric field can strongly quench luminance of the QD film in the absence of mobile charge,[21] with 90% max PL suppression achieved at an electric field of 4MV/cm. Few groups have leveraged QD PL quenching in biological[22] or display[23] applications. For instance, Rowland *et al.*, adopting the capacitor structure, utilized QD emission to track action potential profile of a firing neuron.[22] More recently, Salihoglu *et al.* demonstrated a color-variable PL display that used QD PL quenching to dim selective pixels by directly injecting electric charges into QDs via a graphene layer.[23] The ultimate degree of PL quenching achieved in previous studies is summarized in Table S1. Regardless of the cause for the quenching limit, the maximum PL reduction achieved has been 90% (corresponding to the contrast ratio of 10:1) with response time in milliseconds at best. This is still far below the levels required for modern displays, which typically have a contrast ratio of 1000:1 or higher for vivid pixel-to-pixel contrast.

The observed challenges of limited quenching and slow PL response are both likely associated with screening of the applied electric field by photo-generated, field-ionized, or electrically injected charges.

These charges can reside in delocalized band-edge states, defect states on the nanocrystal, or electronic states in the surrounding matrix. We note that the application of electric field is often also associated with QD charging, even if charging is not the intended outcome. At zero field, a number of studies on single QD blinking[16,24] shows that a photo-generated carrier can be spontaneously ejected to a localized state residing on the QD surface or the surrounding matrix. On an ensemble level, the photo-generated charge could contribute to the screening of the externally applied field,[17,25] which limits the ultimate degree of achievable field-induced quenching.

In this work, we use a QD-LED structure that operates efficiently in forward bias, hence optimized for minimum deleterious charge buildup next to the luminescent QD layer. The device is reverse biased to minimize charge injection from the electrodes into the device. When optically excited to generate the QD PL, the photogenerated charge is rapidly dissociated in the reverse-bias electric field, leading to efficient reduction of the QD PL. Through steady-state and transient PL measurements using excitation light of wavelength λ = 405 nm, we observe up to 99.5% QD PL suppression at 3.3MV/cm, corresponding to a contrast ratio of 200:1. The device also demonstrates reversible modulation with a response time of 300 ns and no significant degradation when continuously operated under millisecond pulsed bias. To our knowledge, this is the best performance reported to date for an electrically-controlled PL device based on colloidal nanocrystals. Using excitation wavelength dependent and time-resolved PL analysis, we find that the high degree of quenching is achieved by QCSE, which is followed by QD exciton dissociation under the reverse-bias electric field.

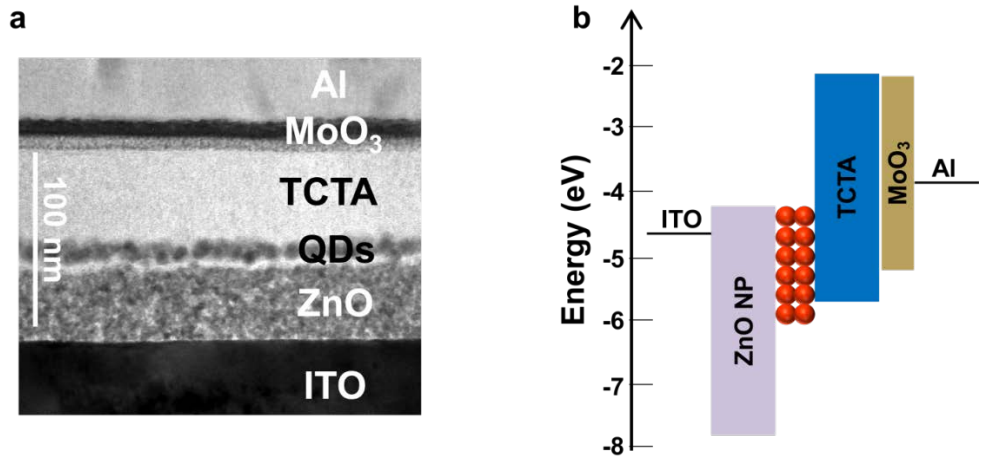

**Figure 1.** (a) Cross-sectional SEM image of the QD-LED prepared by focused ion beam showing each layer of the device. (b) Flat band energy diagram of constituent layers on an energy scale referenced to the vacuum level.

**Results and Discussion**

Our device is comprised of 10 nm (~1 monolayer) thick layer of CdSe/ZnCdS core-shell QDs sandwiched between a ZnO electron transport layer (ETL) and a hole transport layer (HTL) of tris(4-carbazoyl-9-ylphenyl)amine (TCTA). Thin film of Aluminum (Al) with molybdenum oxide ($MoO_3$) underlayer and a film of indium tin oxide (ITO) are used as anode and cathode layers, respectively. The multilayer device is visualized in the cross-sectional scanning electron microscope image (Figure 1a), which shows the thickness of each layer. As shown in Figure S1, the synthesized CdSe/ZnCdS QDs exhibit an emission peaked at $\lambda = 625$ nm with a full width half maximum (FWHM) of 30 nm. The photoluminescence quantum yield (PLQY) of the QD solution and the thin film are measured to be 85% and 75%, respectively. The electroluminescence (EL) performance under forward bias of the QD-LED is also shown in Figure S2. The device exhibits an EL emission peaked at $\lambda = 635$ nm with peak external quantum efficiency (EQE) of 15.9% at a driving voltage of 2.9 V. Assuming a light outcoupling efficiency of between 20% and 25%,[4] the internal quantum efficiency (IQE) of the device is between 65% and 80%, nearly matching the PLQY, and therefore in a forward-biased device the QD film is minimally quenched by the surrounding charge transport layers. The energy band diagram of the device is shown in Figure 1b.

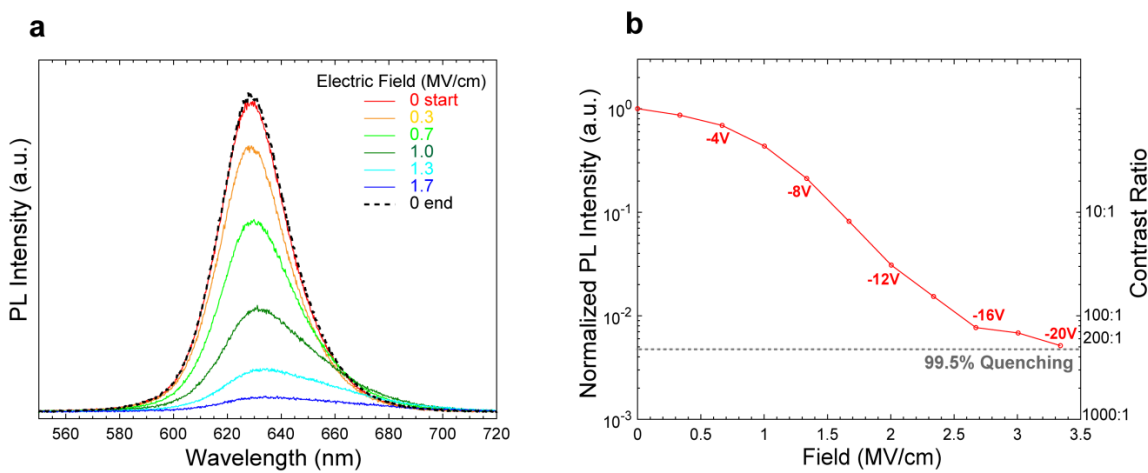

**Figure 2.** (a) Steady state PL spectra of a QD-LED with 10 nm thick QD film, optically excited with $\lambda = 405$ nm light. Drop in PL intensity and a red shift in peak PL wavelength is observed with increasing reverse-biased electric field. The PL fully recovers after the field is removed. (b) Integrated PL intensity (normalized) as a function of applied electric field shows the maximum achievable quenching of 99.5% at 3.3 MV/cm. The corresponding reverse bias voltage is indicated in the plot.

To study the effect of the applied electric field on the QD PL in this structure, we first perform steady-state measurements by applying reverse bias to the device that is optically excited by $\lambda = 405$ nm light, and record the QD PL intensity and spectrum. In reverse-biased QD-LED, the electric field is dropped across the QD layer, but without external charge injection from the device electrodes. As no charge is injected into the device, any observed quenching is related to the applied electric field. We apply 2 ms duration square voltage pulses at 100 Hz repetition rate with amplitude from 0 V up to -20 V, corresponding to electric fields of 0 to 3.3MV/cm. The square waveform is selected to minimize the potential permanent electric field induced damage on the QD film. Electric field strength in the QD film is estimated by dividing the applied reverse bias by the total thickness of the QD and HTL films, which have similar dielectric constants.[26] The voltage drop across the ZnO film is assumed to be minimal due to its high electron mobility.[27] We note that while we report here the value of the applied macroscopic field, the internal field inside the QD is lower due to the dielectric screening from QD surfaces. The precise value of the internal field is difficult to calculate due to the anisotropic, heterogeneous, and partially disordered local environment. As shown in Figure 2a, the steady-state QD PL spectra are plotted with increasing electric field strength. The PL intensity is fully recovered after the electric field is removed, as shown by the good match in Figure 2a between the solid and dashed spectra that are both obtained at 0V applied bias. In addition to the PL intensity decrease, the PL spectra also display a red-shift, as expected from QCSE. The integrated PL intensity (normalized to the intensity at zero bias) is shown in Figure 2b. The device exhibits a maximum 99.5% PL quenching under at 3.3 MV/cm, corresponding to a contrast ratio of 200:1. No further improvement of PL quenching is observed beyond -20 V. CCD camera image of the QD-LED pixel under the microscope starts to show residual emission from isolated spots at -24 V, as illustrated in Figure S3, which is suggestive of the dielectric breakdown at the very high reverse bias.[28,29]

To confirm that the increasing reverse bias leads to an increasing electric field and no charge injection, we measure the electroabsorption spectrum of QDs in the device under millisecond voltage pulses. Figure S4a shows the acquired spectra, with bleached absorption at the first exciton peak around $\lambda = 620$nm and induced absorption on the lower energy side. We can estimate the total oscillator strengths at each field by integrating the electroabsorption spectra over the emission range from 1.77 eV ($\lambda = 700$ nm) to 2.70 eV ($\lambda = 460$ nm), as shown in Figure S4b. Due to oscillator strength sum rules,[30] this integral is conserved if state-filling by charge injection does not occur. While the range of energies in the measurement is limited by the laser, the relative flatness of the oscillator strength integral from 0 V to -20 V, as compared to the sudden increase at -24 V, suggests that charge injection is minimal until the dielectric breakdown.

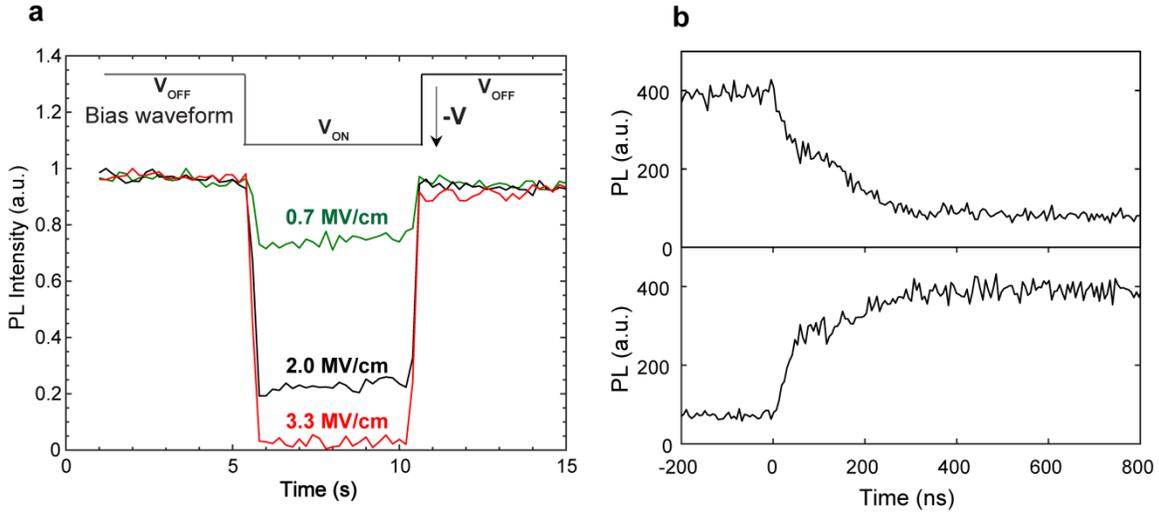

**Figure 3.** (a) Time traces of the device as a function of the applied electric field showing the reversible modulation of PL intensity. Inset: Applied bias waveform. (b) PL intensity measured at the start (top) and the end (top) of a -10V pulse for a device with a 10 nm thick QD layer. The fast response time and recovery time are on the order of three hundred nanoseconds.

To evaluate whether the applied electric field is screened by charges, we apply an extended bias pulse to monitor the temporal evolution of QD PL. Figure 3a displays the PL time traces of the device under 5s square voltage pulses from 0 to -20 V. By overlaying the different traces, we can compare both the maximum PL reduction and the quenching dynamics. In addition to the expected nearly instantaneous PL intensity reduction at the beginning of the electric field application, we also observe a slow recovery in PL intensity while the bias is held constant. We note this PL transient dynamics is more prominent at the lower field (0.7 MV/cm) but is almost completely suppressed at -20 V (3.3 MV/cm). The fact that these dynamics occur on a time scale of hundreds of milliseconds to seconds are consistent with charge buildup in our devices at lower fields. Since we do not have significant charge injection under reverse bias operation, this could be a result of ionization of photo-generated charges. Either the extracted charge or the opposite charge that is left on the QD screens the external electric field; thereby reduces the amount of field-induced quenching on the nearby QDs. In addition to the constant-bias dynamics, the time response of the device at the start and end of the voltage pulse is measured using an avalanche photodiode (APD), as shown in Figure 3b. The device exhibits a fast response time of about 100ns and 300ns to reach 50% and 90% of the maximum quenching, respectively.

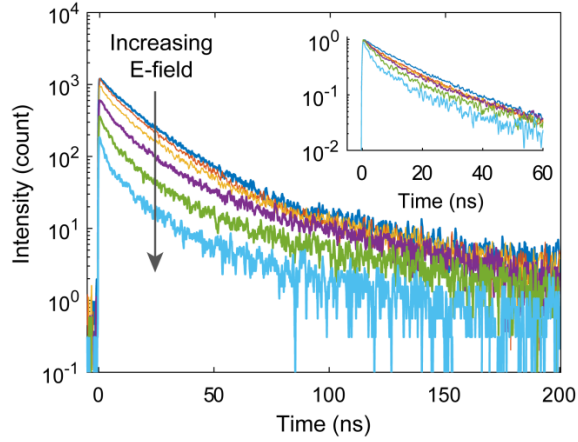

**Figure 4.** Time-resolved PL decays of a QD-LED with 10 nm thick QD layer reverse biased with increasing electric field from 0 to 1.7 MV/cm. Inset: normalized decay traces. The samples are excited using a λ = 405 nm pulsed laser.

In order to identify the mechanisms behind field-induced PL quenching, we investigate exciton decay dynamics at different applied electric fields using time-resolved photoluminescence (TRPL) measurements. Figure 4 shows QD PL time transients under increasing electric fields from 0 to 1.7 MV/cm. It is noted that the TRPL measurement is limited by insufficient PL signals beyond -10 V (1.7 MV/cm) due to the efficient PL quenching. The normalized traces (in the inset) show that the PL decay becomes faster with increasing field. All decay curves are fitted to a bi-exponential function. The resulting decay lifetimes are extracted and shown in Table 1. Two decay lifetimes, $\tau_1$ and $\tau_2$, exhibit distinct trends as the electric field is increased from 0 to 1.7 MV/cm. While $\tau_1$ shows a marked decrease from 8.58 ns to 2.99 ns, $\tau_2$ displays a small change in the range of 20 ns to 23 ns. This trend differs from multiple earlier investigations, in which the PL decay lifetimes remained constant with electric field.[10,21,22] To explain this difference, we note that the higher luminescence efficiency of QDs used in this work would make our QD luminescence lifetimes particularly susceptible to the introduction of a new non-radiative pathway, such as the field-induced exciton dissociation. Also, the primary difference between our device and those in the earlier reports is the existence of the charge transport interfaces sandwiching the 1 to 2 monolayers of the QD film, facilitating rapid transport of photogenerated charges away from the QD.

**Table 1. Decay lifetimes as a function of applied electric field**

| E-field (MV/cm) | $\tau_1$ (ns) | $\tau_2$ (ns) | $A_1$ | $A_2$ | Relative Fraction | |
|---|---|---|---|---|---|---|
| 0 | 8.58 | 22.61 | 618 | 647 | 0.489 | 0.511 |
| 0.3 | 6.08 | 20.09 | 518 | 720 | 0.419 | 0.581 |
| 0.7 | 5.57 | 21.16 | 489 | 502 | 0.493 | 0.507 |
| 1 | 5.09 | 22.63 | 375 | 267 | 0.584 | 0.416 |
| 1.3 | 4.97 | 23.08 | 242 | 125 | 0.659 | 0.341 |
| 1.7 | 2.99 | 20.50 | 124 | 63 | 0.664 | 0.336 |

The QD film in the reverse-biased QD-LED can act as a photoactive layer, where upon optical excitation of a QD the excited carrier can be swept out of the QD into the neighboring charge transport layer. The photogenerated carriers in the core of the QD need to tunnel through the wider bandgap QD shell material to reach the charge transport layers. This charge extraction process is easier for hot carriers, which are generated by photons whose energy is significantly larger than the QD core bandgap energy. Mashford *et al.* have previously demonstrated, by using Kelvin probe measurements of QD films on top of ZnO nanoparticle films, that photoelectrons are extracted from CdSe/CdS core-shell QDs and transferred to ZnO when excited by high optical excitation energy (above QD bandgap).[31] Other reports have shown that the relative rate of charge extraction versus relaxation to the band-edge is dependent on the excitation energy in QDs,[32,33] and this can lead to an excitation energy dependence of the PL if the trapped charges are ultimately extracted by the electrodes or if they recombine through non-radiative channels.[34,35] Data in Figure 5 is consistent with this model, as we measure the steady-state PL quenching of the QD film that is excited with different excitation wavelengths, swept from $\lambda = 460$ nm to $\lambda = 560$ nm. No significant change in the amount of PL quenching for different excitation wavelengths is observed until the applied bias reaches -8 V (1.3 MV/cm). As we keep increasing the electric field, the device shows a sharper drop in PL with higher photon energy excitations. At -20 V bias (3.3 MV/cm) quenching of QD PL generated by $\lambda = 460$ nm excitation is a factor of 3.5 more efficient than the quenching of QD PL generated by $\lambda = 560$ nm excitation. The dependence of the PL quenching efficiency on the energy of the excitation light is consistent with field-induced extraction of photogenerated charges out of the QDs, process which is more efficient when hot carriers are generated. Given the amorphous structure of QD interfaces with neighboring charge transport layers, it is likely that the energy barrier for charge extraction is different locally for different individual QDs. As a result, even at higher electric fields charge extraction from some of the individual QDs could be impeded. This subset of QDs (which we refer to as QD subset 2) would have an unchanging PL lifetime $\tau_2$ (21.7 ± 1 ns) for different bias conditions, just as observed in our time-resolved fits of QD PL at different biases (Table 1). Our measured $\tau_2$ lifetime is consistent with previous observations, which have shown 20 ns to 30 ns radiative lifetime of neutral excitons in PL-efficient CdSe QDs.[16,36] The other subset of QDs (which we refer to as QD subset 1) manifests a reduced effective PL lifetime, $\tau_1$ in Table 1, as the field-induced charge extraction process dissociates the excitons on these QDs, introducing a new non-radiative process that is more-rapid than the radiative recombination. The fractional areas under the decay curve, $A_1$ and $A_2$, that have lifetimes $\tau_1$ and $\tau_2$, respectively, represent the relative contributions to the measured QD PL from the two subsets of luminescing QDs. Both $A_1$ and $A_2$ decrease with the increasing electric field, with $A_2$ decreasing more rapidly.

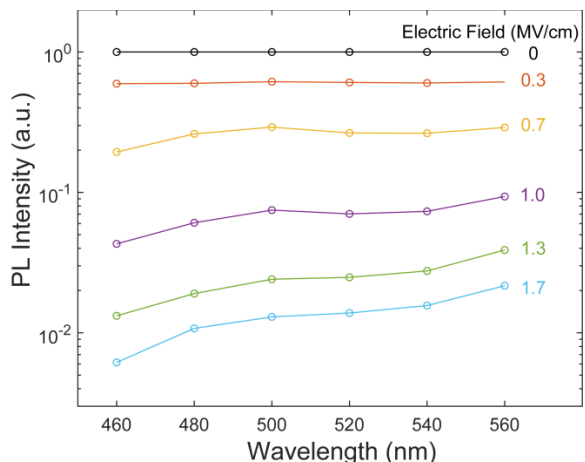

**Figure 5**. Integrated PL intensity (normalized) of the device as a function of the excitation wavelength (from $\lambda = 460$ nm to $\lambda = 560$ nm) with increasing applied electric field.

The two subsets of QDs, the subset 1 perturbed by the applied electric field and the subset 2 not perturbed, are also evident in the PL spectra taken at different electric fields, shown in Figure 2a. The measured PL spectra can be fitted by using two Voigt functions, as shown in figure S5a for QD PL spectra at fields of 0 and 1.7 MV/cm. Figure S5b plots the peak wavelength of the two Voigts, labeled as peak 1 (low energy, high wavelength peak) and peak 2 (high energy, low wavelength peak), for different electric fields. With increasing electric field, it is evident that peak 1 is strongly red-shifted (from $\lambda = 635$nm to $\lambda = 653$nm), while peak 2 is relatively steady with the red-shift of only 2 nm. Given that the PL peak wavelength of our QDs in solution is $\lambda = 625$ nm, we associate peak 2 with the QD subset 2, whose TRPL decay lifetime is $\tau_2$. The red shift of peak 1 is consistent with the QCSE, and should be accompanied by the reduction in TRPL decay lifetimes, just as we see for QD subset 1 in our TRPL traces. In addition, the relative fraction of peak 2 drops from 53% at 0 MV/cm to 25% at 1.7 MV/cm (Figure S5c), which is consistent with the TRPL lifetime results for QD subset 2 in Table 1.

Using data of Table 1 we can then estimate the expected QD PL quenching at different bias conditions. First we need to estimate the fraction of excitons that are responsible for the QD PL signal at 0V applied bias, since at the no-bias condition we see from Table 1 that about half of the excitons that luminesce are from QDs in subset 1 and half are in subset 2, with $\tau_2/\tau_1 = 2.6$. This implies that although the two subsets contribute equally to the observed PL, subset 1 has 2.6 times more excitons, with one in 2.6 of them luminescing, and the others being quenched. The total number of luminescent QD excitons then appears to be 2 out of 3.6. However, this should further be modified to account for 75% PL efficiency of QDs when in a thin film form, hence yielding that 2 out of 4.8 QDs in a QD-LED structure (or 42% of optically excited QDs) are providing for the QD PL signal.

At the electric field of 1.7 MV/cm (-10V applied bias), about 2/3 of PL emission is due to the QD subset 1, which experiences the applied electric field and can undergo exciton breakup by injecting charge into the neighboring transport layers. The few sublet 1 excitons that do recombine and generate PL have to do so rapidly, before the field rips them apart, so their observed PL lifetime is $\tau_1 = 2.99$ ns. Without the neighboring charge transport layers, the QD excitons might not be dissociated. They could undergo the QCSE redshift and might eventually radiatively recombine, but with a significantly longer luminescence lifetime, as the electron and hole that comprise the exciton would have a reduced wavefunction overlap.

The data in Figure 2b indicates that at 1.7 MV/cm there is a 12.5-fold decrease in the observed PL signal as compared to the 0V bias condition. This would be the case if the radiative lifetime of subgroup 1 is extended to $\tau_{1\text{-RAD}} = 99$ ns due to QCSE. As we observe $\tau_1 = 2.99$ ns at 1.7 MV/cm, then $\tau_{1\text{-RAD}}/\tau_1 = 33$ ns. This implies that in QD subgroup 1, 33-times more excitons are dissociated by the applied field than the ones that provide for the PL signal. In addition, 1/3 of the PL signal is from QDs in subgroup 2, half as much as from subgroup 1 PL. In aggregate, this implies that from 33.5 excitons, 1.5 excitons will luminesce. Accounting again for the 75% PL efficiency of QDs when in a thin film form we get that out of 45 optically excited QDs in a QD-LED structure 1.5 (or 3.3% of optically excited QDs) will provide for the QD PL signal. This is 12.5-fold less luminescence (42% / 3.3% = 12.5) than we predicted above for the no-bias condition.

Consistency of our analysis depends on the extension of $\tau_{1\text{-RAD}}$ from 21.7ns under no bias condition to 99 ns under an electric field of 1.7 MV/cm due to QCSE. Such lifetime extension was previously proposed by Bozyigit *et al.*, where they estimated that at 1.7 V/cm, the QD PL radiative lifetime could increase by between 1.6 and 5-times for a 4nm core-diameter CdSe QD coated with 2.1nm thick ZnS and CdS shells respectively.[21] If only QD subgroup 1 experiences QCSE, this would imply that the radiative lifetime of subgroup 1 QDs would be 1.6 to 5 times longer than that of subgroup 2 QDs, hence 35 ns to 109 ns, which is consistent with our assumption of $\tau_{1\text{-RAD}} = 99$ ns.

**Conclusion**

In summary, we demonstrate that QD-LEDs can be used as an efficient voltage-controlled optical down-converter when operating under reverse bias. We use field-induced QD luminescence quenching in our favor, and show efficient PL modulation of up to 99.5% (200:1 contrast ratio) under 3 MV/cm with three hundred nanosecond response time, giving us the best performing electrically-controlled QD PL device to date. Using optical spectroscopy we further show that a large electric field can strongly modify carrier dynamics in a nanostructure, and the stable and repeatable quenching is accomplished by QCSE and field-induced exciton dissociation. Quantifying the time-resolved spectroscopic data, as well as the observed spectral shifts due to the quantum-confined Stark effect, we are able to numerically match the observed PL quenching, providing a comprehensive physical mechanism for the observations. The results presented open up additional research directions as similar device structures could be used in electro-optic modulators or electrically-tuned PL display technology. While our work demonstrated the validity of using a reverse-biased QD-LED to access the high field regime for exciton control, there is still much room for optimization in terms of both material and device architecture, and we anticipate that the results reported herein will facilitate the rational design and development of QD materials that can be integrated into nanostructured devices.

**Methods**

**Synthesis of Colloidal Quantum Dots (QDs)**

*Synthesis of CdSe Seed Cores:* 262.5 mmol of cadmium acetate was dissolved in 3.826 mol of tri-n-octylphosphine at 100°C in a 3 L 3-neck round-bottom flask and then dried and degassed for one hour. 4.655 mol of trioctylphosphine oxide and 599.16 mmol of octadecylphosphonic acid were added to a 5 L stainless steel reactor and dried and degassed at 140°C for one hour. After degassing, the Cd solution was added to the reactor containing the oxide/acid and the mixture was heated to 310°C under nitrogen. Once the temperature reached 310°C, the heating mantle is removed from the reactor and 731 mL of 1.5 M diisobutylphosphine selenide (DIBP-Se) (900.2 mmol Se) in 1-Dodecyl-2-pyrrolidinone (NDP) was then rapidly injected. The reactor was then immediately submerged in partially frozen (via liquid nitrogen) squalane bath rapidly reducing the temperature of the reaction to below 100°C. The first absorption peak of the nanocrystals was at $\lambda = 480$ nm. The CdSe cores were precipitated out of the growth solution inside a nitrogen atmosphere glovebox by adding a 3:1 mixture of methanol and isopropanol. After removal of the methanol/isopropanol mixture, the isolated cores were dissolved in hexane and used to make core-shell materials.

*Growth of CdSe cores:* A 1 L glass reactor was charged with 320 mL of 1-octadecene (ODE) and degassed at 120°C for 15 minutes under vacuum. The reactor was then backfilled with N2 and the temperature set to 60°C. 120 mL of the CdSe seed core above was injected into the reactor and the hexanes were removed under reduced pressure until the vacuum gauge reading was < 500 mTorr. The temperature of the reaction mixture was then set to 240°C. Meanwhile, two 50 mL syringes were loaded with 80 mL of cadmium oleate in TOP (0.5 M conc.) solution and another two syringes were loaded with 80 mL of di-iso-butylphosphine selenide (DiBP-Se) in TOP (0.5 M conc.). Once the reaction mixture reached 240°C, the Cd oleate and DiBP-Se solutions were infused into the reactor at a rate of 35 mL/hr. The 1st excitonic absorption feature of the CdSe cores was monitored during infusion and the reaction was stopped at ~60 minutes when the absorption feature was peaked at $\lambda = 569$ nm. The resulting CdSe cores were then ready for use as is in this growth solution for overcoating.

*Synthesis of CdSe/ZnCdS Core-Shell Nanocrystals*: 115 mL of the CdSe core above with a first absorbance peak at $\lambda = 569$ nm was mixed in a 1 L reaction vessel with 1-octadecene (45 mL), and Zn(Oleate) (0.5 M in TOP, 26 mL). The reaction vessel was heated to 120°C and vacuum was applied for 15 min. The reaction vessel was then back-filled with nitrogen and heated to 310°C. The temperature was ramped, between 1°C/5 seconds and 1°C/15 seconds. Once the vessel reached 300°C, octanethiol (11.4 mL) was swiftly injected and a timer started. Once the timer reached 6 min., one syringe containing zinc oleate (0.5 M in TOP, 50 mL) and cadmium oleate (1 M in TOP, 41 mL), and another syringe containing octanethiol (42.2 mL) were swiftly injected. Once the timer reached 40 min., the heating mantle was dropped and the reaction cooled by subjecting the vessel to a cool air flow. The final material was precipitated via the addition of butanol and methanol (4:1 ratio), centrifuged at 3000 RCF for 5 min, and the pellet redispersed into hexanes. The sample is then precipitated once more via the addition of butanol and methanol (3:1 ratio), centrifuged, and dispersed into toluene for storage.

*Characterization:* Absorbance and PL spectra of QD solution were obtained by a Cary 5000 UV−Vis−NIR infrared spectrometer and a Fluoromax-3 spectrofluorometer (Horiba Jobin Jvon) respectively. The absolute PL quantum yield (PLQY) of QD films were measured using an integrated sphere in air with a $\lambda = 405$ nm laser excitation. TEM of QDs was taken using a FEI Tecnai G2 microscope. Structural characterization of the QD-LED cross-section was carried using a FEI Helios Nanolab 600 duel beam FIB/SEM.

**Synthesis of ZnO Nanoparticles**

The ZnO nanoparticles synthesis was adapted from a previously published protocol[31]. In a typical synthesis, 1.5 g of zinc acetate dihydrate was dissolved in 100 mL of 2-methoxyethanol. The solution was stirred for 3 h under room temperature. 25 g of tetramethylammonium hydroxide (TMAH) dissolved in 10 mL of 2-methoxyethanol was prepared and added to the zinc acetate solution. Afterwards, the solution was washed with toluene and hexane then re-dispersed in a mixture of methanol and isopropanol. The solution was filtered with PTFE 0.2µm filters and stored in a freezer.

**Device Fabrication**

QD solutions are purified three times by precipitation with butanol and methanol and then redispersion in octane. Prior to device fabrication, pre-patterned indium-tin-oxide (ITO) substrates (obtained from thin film devices) were cleaned by sonication in detergent, DI water, acetone (5 min each), followed by boiling in isopropanol twice for 5 min. Oxygen plasma was subsequently used to treat the substrate for 10 min. ZnO nanoparticles dissolved in a mixture of methanol and isopropanol were then spin-cast onto the ITO substrates at 2000 rpm for 60 s and annealed at 100°C for 45 min in a nitrogen-filled glovebox. QD solution was spun at 2000 rpm for 60 s. QD film thickness was controlled by changing the solution concentration. A 50 nm thick HTL of tris(4-carbazoyl-9-ylphenyl)amine (TCTA; Lumtec > 97%), a 10 nm thick layer of molybdenum oxide ($MoO_3$; Alfa Aesar 99.9995%), and a 100 nm thick Al top electrode were thermally evaporated through a shadow mask at a base pressure of $10^{-6}$ Torr.

**Device Characterization**

*QD electroluminescence characterization*: The QD-LED performance under forward bias was characterized inside a nitrogen-filled glovebox. Electroluminescence (EL) spectrum was measured using an Ocean Optics SD2000 fiber-coupled spectrometer. Current-density-voltage (J-V) characteristics were recorded using a computer-controlled Keithley 2636A source meter. A calibrated Newport 818-UV silicon photodiode was used to measure the optical power output. Luminance and EQE were then calculated by assuming Lambertian emission profile and accounting for the wavelength dependence (weighted by the device's EL spectrum) of the photodiode's responsivity[37].

*Steady state PL:* All of steady-state PL measurements are performed in air on devices that were encapsulated in a nitrogen glovebox prior to the measurements. A $\lambda = 405$ nm pulsed laser (PicoQuant) with 80MHz repetition rate is used as a quasi-CW excitation light source. For measurements where the excitation is synchronized to the bias, the laser is electrically gated, with the leading and trailing edge of the laser pulse shifted by 5 μs toward the middle of the bias pulse, so that the measured PL is not affected by any electrical transients. The laser is incident on the device at approximately 20 degrees from normal through the glass/ITO, while the PL is collected by a pair of relay lens into a fiber bundle and delivered to an Acton 2300i imaging spectrometer equipped with a 300 l/mm grating and a Pixis 100b cooled CCD. A long-pass filter between the relay lens filters out the scattered $\lambda = 405$ nm light. For PL time trace measurements, the spectra are collected continuously. CCD images are acquired using a QImaging EXi camera, with the device viewed under a 10x objective lens on a microscope and illuminated with a $\lambda = 405$ nm LED.

*Excitation wavelength dependent PL:* Excitation wavelength dependent photoluminescence measurements are taken using an NKT Photonics SuperK laser with an AOTF as the excitation source. The same optical chopping and spectrometer detection scheme is used as for the $\lambda = 405$nm excitation photoluminescence quenching measurements. At each bias, excitation wavelength is changed every 50ms. The entire list of excitation wavelengths is cycled through and repeated an integer number of times until the desired signal-to-noise ratio is reached.

*Time-resolved PL:* Time-resolved photoluminescence data are taken using a Si SPAD detector and a PicoHarp 300 Time-Correlated Single Photon Counting module. The samples are excited using a $\lambda = 405$ nm PicoQuant pulsed laser at a repetition rate of 2.5MHz. The instrument response of the system at the detected wavelength is measured to have a FWHM of less than 0.5 ns using the Allura Red dye. The QD-LEDs are driven using square pulses of 1ms duration and 10 ms period to limit photo-charging effects. Both the laser and the SPAD are gated such that the laser pulses are produced and photons are collected only when the bias pulses are high.

*Transient absorption experiments:* A SuperK laser with an acouto-optic tunable filter (AOTF) is used as the light source. A collimated s-polarized beam reflects off the Al electrode of the QD-LEDs and is collected by a home-made voltage-tuned balanced photodiode. The differential absorption laser power used is <1uW/mm$^2$ and much smaller than the $\lambda = 405$ nm photo-excitation laser. A 100 point spectrum can be taken in 2s by sweeping the AOTF from $\lambda = 460$ nm to $\lambda = 700$ nm.

**Corresponding Author**

* bulovic@mit.edu

**Acknowledgements**

The project was initiated under the Center for Excitonics, an Energy Frontier Research Center funded by the US Department of Energy, Office of Science, and Office of Basic Energy Sciences under Award Number DE-SC0001088 (MIT). The project was completed with support of the Material & Device Research Institute of the Samsung Advanced Institute of Technology (SAIT). This project made use of the MRSEC Shared Experimental Facilities at MIT, supported by the National Science Foundation under award number DMR-14-19807. The authors thank Robert Nick for kindly providing colloidal quantum dot solution, and thank Michel Nasilowski for providing ZnO nanoparticle solution.

**Author Contributions**

‡These authors contributed equally to this work. All authors discussed the results and contributed to editing the manuscript.

**Notes**

The authors declare no competing financial interest.
**Supporting Information**

The supporting information is available free of charge on ACS Publication website

Additional information such as the absorption and photoluminescence spectra of the quantum dots in solution, device performance under forward bias, CCD camera images of device under microscope, electroabsorption spectrum of QD film in device under reverse bias, and fitting of PL spectrum.

# Supporting Information

Voltage-Controlled Reversible Modulation of Colloidal Quantum Dot Thin Film Photoluminescence

*Sihan Xie†, Han Zhu†, Melissa Li, Vladimir Bulović\**

Research Laboratory of Electronics, Massachusetts Institute of Technology, Cambridge, Massachusetts 02139, United States

†These authors contributed equally to this work.
\*Corresponding author. Email: bulovic@mit.edu

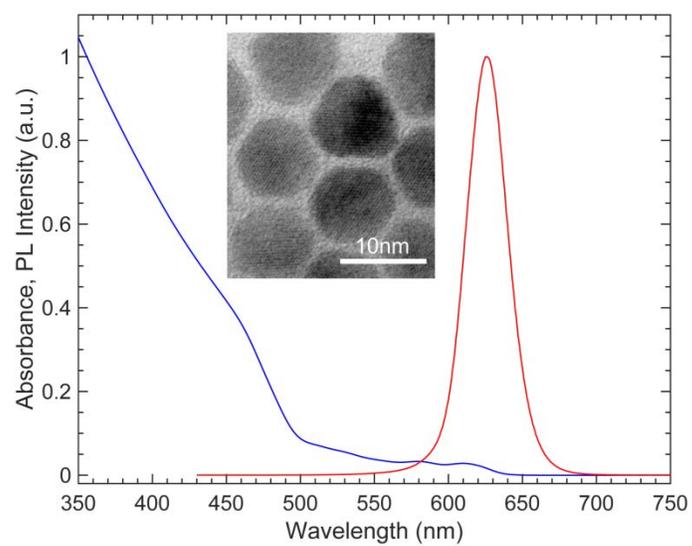

**Figure S1.** Absorption (blue line) and photoluminescence (red line) of red CdSe-ZnCdS QDs used in the device. Inset: TEM image of QDs. Scale bar is 10 nm.

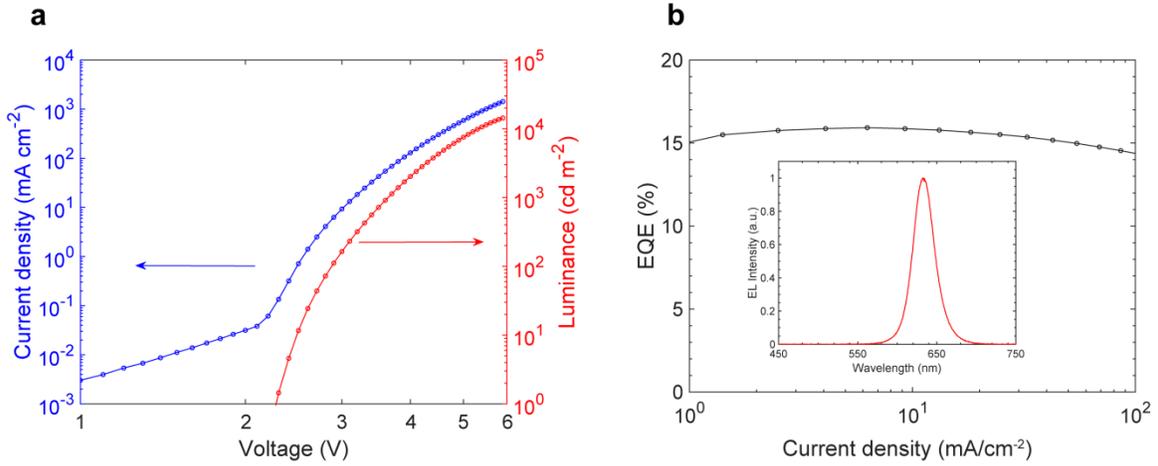

**Figure S2.** QD-LED (with 10 nm thick QD layer) performance under forward bias. (a) current density-voltage-luminance characteristics (b) EQE versus current density. Inset: electroluminescence spectrum

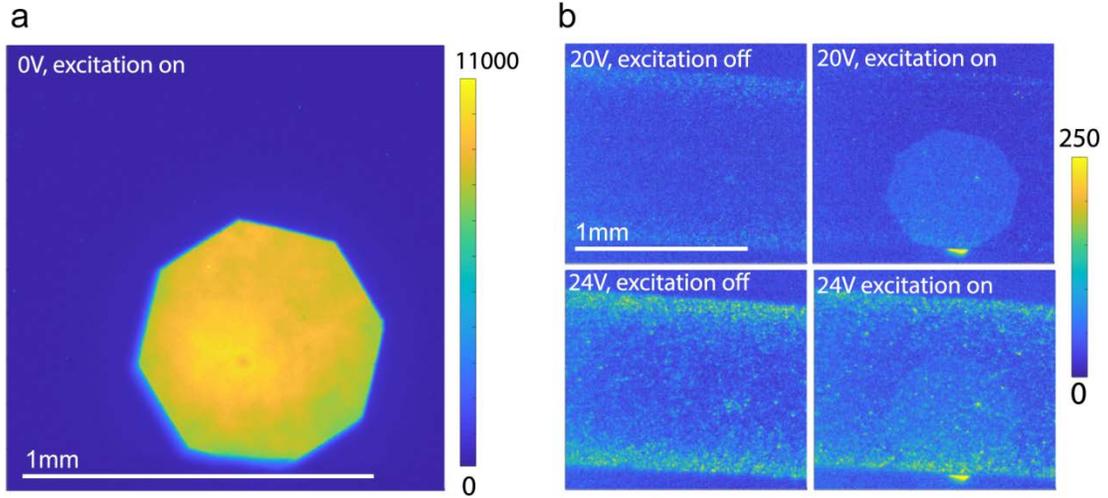

**Figure S3.** CCD camera images of QD-LED (with 10 nm thick QD layer) under microscope illuminated with a λ = 405 nm LED under a) 0 V and b) -20 V and -24 V reverse bias. Illumination is masked outside of the octagonal iris. The faint rectangular shape in each sub-figure of (b) outlines the active region of the device. The residual luminance observed at -24 V is due to EL resulting from dielectric breakdown.

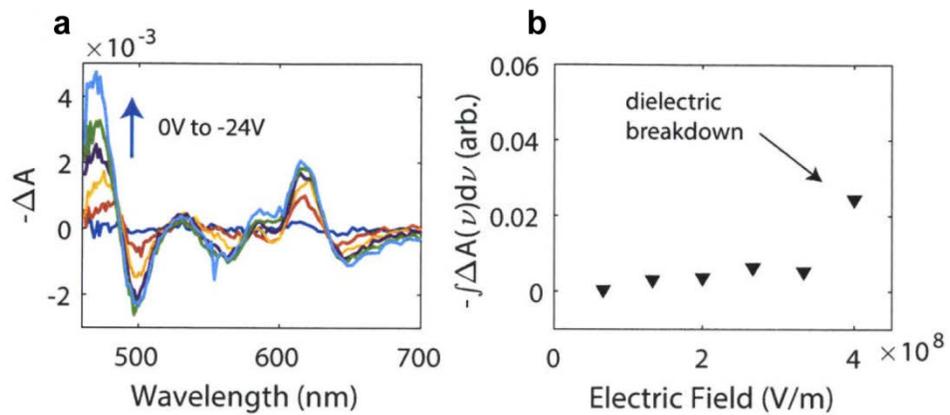

**Figure S4**. a) Electroabsorption spectrum of a monolayer thick QD film in a QD-LED under reverse bias. (b) Integral of the electroabsorption spectrum from 1.77eV (λ = 700nm) to 2.70eV (λ = 460nm).

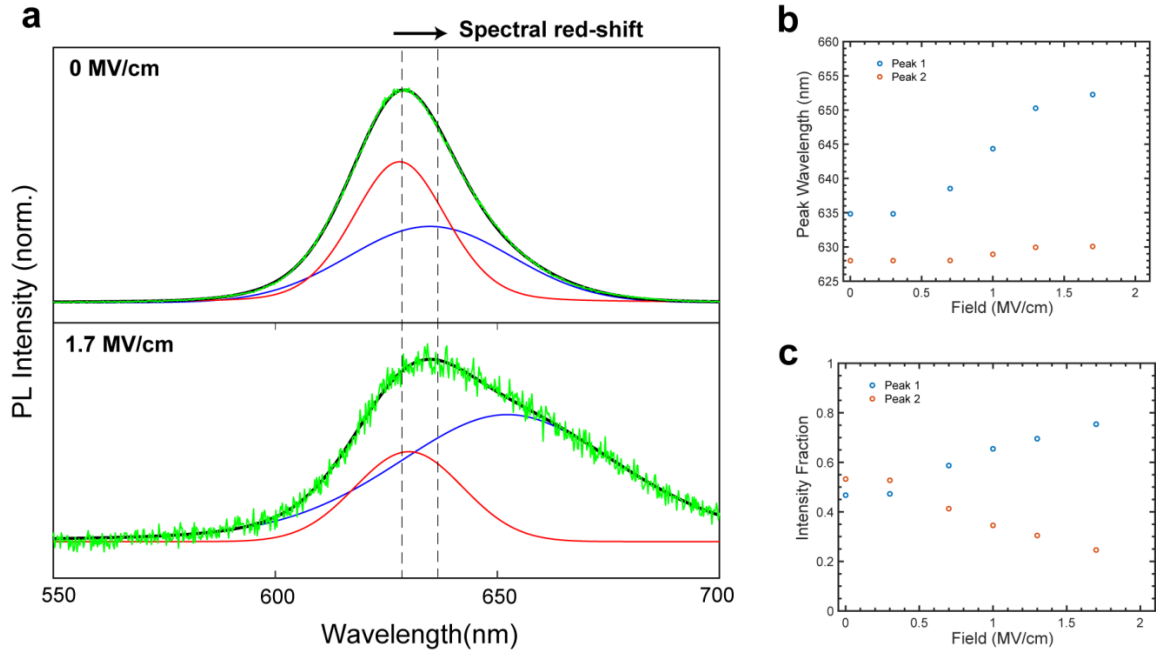

**Figure S5**. a) PL spectrum fit for a device with 10 nm thick QD layer is shown in a) under no field, and b) under the field of 1.7 MV/cm. The raw spectrum data, model fit, and the two Voigt fit functions are shown in black, green, red, and blue, respectively. The peak wavelength and intensity fraction of the two peaks are plotted as a function of field from 0 to 1.7MV/cm in b) and c).

**Table S1. Summary of previously reported voltage-controlled photoluminescence quenching of quantum dot ensembles**

| Reference | Year | QD Material | Maximum PL Quenching |
|---|---|---|---|
| Korlacki et al.[1] | 2011 | CdSe in copolymer | 90% at 4 MV/cm |
| Bozyigit et al.[2] | 2012 | CdSe/ZnCdS | 60% at 3 MV/cm |
| Bozyigit et al.[3] | 2013 | CdSe/CdS | 90% at 4 MV/cm |
| Rowland et al.[4] | 2015 | CdSe/ZnSe in PMMA | 70% at 0.8 MV/cm |
| Prasai et al.[5] | 2015 | CdSSe on $MoS_2$ | 75% at 2V |
| Moebius et al.[6] | 2016 | CdSe/CdS | 60% at 10V |
| Scott et al.[7] | 2016 | CdSe Nanoplatelets in polymer | 28% at 0.13 MV/cm |
| Rowland et al.[8] | 2017 | CdSe/CdS in PMMA | 50% at 6-8 MV/cm |
| Salihoglu et al.[9] | 2018 | CdSe/ZnS on graphene | 70% at 20V |
| This work | 2020 | CdSe/ZnCdS in QD-LED | 99.5% at 3 MV/cm |